\documentclass[twocolumn]{emulateapj}
\usepackage{amsmath}
\begin{document}
\title{Debris Disks in the Scorpius-Centaurus OB Association Resolved by ALMA}
\author{Jesse Lieman-Sifry\altaffilmark{1}, 
A. Meredith Hughes\altaffilmark{1},
John M. Carpenter\altaffilmark{2,3},
Uma Gorti\altaffilmark{4},
Antonio Hales\altaffilmark{3,5},
Kevin M. Flaherty\altaffilmark{1}
}
\altaffiltext{1}{Department of Astronomy, Van Vleck Observatory, Wesleyan University, 96 Foss Hill Dr., Middletown, CT 06459, USA}
\altaffiltext{2}{Department of Astronomy, California Institute of Technology, MC 249-17, Pasadena, CA 91125, USA}
\altaffiltext{3}{Atacama Large Millimeter/Submillimeter Array, Joint ALMA Observatory, Alonso de C\'{o}rdova 3107, Vitacura 763-0355, Santiago, Chile}
\altaffiltext{4}{SETI Institute, Mountain View, CA, USA; NASA Ames Research Center, Moffett Field, CA, USA}
\altaffiltext{5}{National Radio Astronomy Observatory, 520 Edgemont Road, Charlottesville, Virginia, 22903-2475, USA}

\slugcomment{Accepted for publication in ApJ: June 20, 2016}

\newcommand{\RNum}[1]{\uppercase\expandafter{\romannumeral #1\relax}}

\begin{abstract}
We present a CO(2-1) and 1240\,$\mu$m continuum survey of 23 debris disks with spectral types B9-G1, observed at an angular resolution of 0\farcs5-1'' with the Atacama Large Millimeter/Submillimeter Array (ALMA). The sample was selected for large infrared excess and age $\sim$10\,Myr, to characterize the prevalence of molecular gas emission in young debris disks. We identify three CO-rich debris disks, plus two additional tentative (3$\sigma$) CO detections. Twenty disks were detected in the continuum at the $>3\sigma$ level. For the 12 disks in the sample that are spatially resolved by our observations, we perform an independent analysis of the interferometric continuum visibilities to constrain the basic dust disk geometry, as well as a simultaneous analysis of the visibilities and broad-band spectral energy distribution to constrain the characteristic grain size and disk mass. The gas-rich debris disks exhibit preferentially larger outer radii in their dust disks, and a higher prevalence of characteristic grain sizes smaller than the blowout size. The gas-rich disks do not exhibit preferentially larger dust masses, contrary to expectations for a scenario in which a higher cometary destruction rate would be expected to result in a larger mass of both CO and dust. The three debris disks in our sample with strong CO detections are all around A stars: the conditions in disks around intermediate-mass stars appear to be the most conducive to the survival or formation of CO.
\end{abstract}

\section{Introduction}

The tenuous, dusty debris disks observed around nearby main sequence stars are thought to be signposts of mature planetary systems.  Since the dust lifetimes are short compared to the age of the star, the dust is believed to be second-generation material, resulting from grinding collisions  of Pluto-like planetesimals \citep[see][and references therein]{wya08}.  These collisions may be catalyzed either by the recently-formed Kuiper Belt analogues themselves \citep[``self-stirred," e.g.,][]{ken04,dom03} or stirring by a giant planet (\citealt[][]{mus09}; see also the discussion by \citealt{ken10}).  Debris disks are common, with recent surveys detecting infrared dust excess around 20\% of nearby FGK stars and 24\% of A stars \citep{eir13,thu14}.  Since current sensitivity limits are insufficient to detect a debris disk comparable to that generated by our own Solar System's Kuiper Belt, these fractions are almost certainly an underestimate of the prevalence of debris disk-hosting systems around nearby stars, and present-day samples represent only more dynamically active, scaled-up versions of the Kuiper Belt.  

A primary interest in studies of debris disks around nearby stars has been spatially resolving their surface brightness structure in order to better understand the clues that the more easily observed dust can provide to the structure of the less easily observed underlying planetary system.  The highest-resolution observations of debris disks to date have been achieved by observing scattered light at optical wavelengths, with spectacular surveys revealing a wide variety of structures (warps, eccentric rings, spiral arms, etc.), many of which have been linked to the presence of planetary systems \citep[e.g.][]{gol06,sch14,sta14,sou14}.  Surveys of structure revealed in thermal emission have also gained traction in recent years, particularly with the advent of sensitive infrared instruments like {\it Herschel} \citep[e.g.,][]{boo13,mor13,paw14}.  A multiwavelength approach is necessary to understand the physical mechanisms underlying the observed disk structure, since observations at different wavelengths probe different populations of dust grains that are affected differently by disk-shaping mechanisms like radiation pressure, gas drag from interstellar medium (ISM) material, and gravitational interactions with unseen planets.  In particular, the longest-wavelength observations probe large dust grains that are not significantly influenced by stellar radiation pressure and ISM interactions that shape disks at optical and near-IR wavelengths \citep[e.g.,][]{wya08,man09,deb09}.

Millimeter-wavelength observations of debris disk structure therefore fill an important niche in studies of planetary system structure and evolution.  Single-dish surveys have provided good sensitivity with limited spatial resolution \citep[e.g.][]{hol98,pan13}.  Until recently, the modest sensitivity of millimeter-wavelength interferometers limited high-resolution measurements of debris disk structure to only the brightest handful of systems, which could typically only be investigated on an individual basis \citep{koe01,wil02,wil11,wil12,man08,cor09,hug11,hug12,pie11,ric13,mac15a,mac15b}.  As continuum sensitivity of interferometers has increased along with their bandwidth, the first small, uniform sample of resolved observations of disks around Solar-type stars was recently gathered \citep{stee15}.  The advent of the ALMA interferometer, with its large collecting area and wide bandwidth, has enabled a significant step forward in the characterization of the surface density structure of debris disks.  Spectacular images of Fomalhaut \citep{bol12}, HD 21997 \citep{moo13}, AU Mic \citep{mac13}, $\beta$ Pictoris \citep{den14}, HD 107146 \citep{ric15}, and 49 Ceti \citep{hug16} have provided the first detailed views of structure at millimeter wavelengths -- including features like asymmetries, rings, gaps, and surface density profiles that both increase and decrease with distance from the central star.  

One major question about the properties and evolution of debris disks is the prevalence of gas-rich debris systems and the origin of that gas.  While this is a question that has been explored sporadically over the past two decades \citep[e.g.][]{zuc95,rob00,den05,red07,hug08}, the sensitivity and resolution of ALMA have enabled exciting new high-resolution characterization of debris disks in molecular emission as well as continuum \citep{kos13,den14,hug16}.  One puzzle that persists is the question of whether the gas in these disks is primordial gas from the protoplanetary disk that has persisted beyond the stage at which the primordial dust disk dissipated (as long as 40\,Myr in the case of 49 Ceti), or whether it is instead second-generation gas resulting from the vaporization of material that had previously been incorporated into the icy mantles of dust grains, Pluto-size planetesimals, or even Mars-size bodies that have undergone a recent collision.  The asymmetric distribution of gas and dust in the $\beta$ Pictoris system is likely the result of either a recent collision between Mars-size bodies or evaporation of icy grain mantles undergoing a higher collision rate at resonant points arising from the influence of an unseen ice giant planet \citep{den14}.  The rest of the disks that have so far been detected have exhibited a more symmetric gas distribution that could be consistent with either a primordial or second-generation scenario; while the gas lifetimes calculated in response to the ionizing radiation produced by the central A star are very short \citep[of order kyr; see discussion in][]{kos13}, and there is spectroscopic evidence that the gas around 49 Cet and $\beta$ Pic is volatile-rich \citep{rob06,rob14}, it is also true that the necessary replenishment rate for the gas is very high, requiring an uncomfortably large vaporization rate of a Hale-Bopp-size body every few minutes \citep{kos13,den14}.  The origin and nature of the gas therefore remains elusive.  

There is an obvious synergy between studies of gas and dust emission in debris disks, out of which the current project was born.  In attempting to ascertain the prevalence of molecular gas emission from continuum-bright debris disks with ages of $\sim$10\,Myr, we surveyed a sample of 23\footnote{The total sample contains 24 disks, but since the time of observation a consensus has emerged that one of the disks in our sample, AK Sco, is a primordial circumstellar disk and not a debris disk.} debris disks in the Sco-Cen star forming region for CO emission.  We cross-matched {\it Spitzer} and {\it IRAS} observations with {\it 2MASS}, {\it WISE}, and {\it Akari}, and the resulting spectral energy distribution (SED) was fitted with a two-component disk model consisting of (1) a stellar photosphere and (2) a modified blackbody representing the dust emission.  This single-blackbody fit to the SED was used only to select a sample of objects with high likelihood of detection at millimeter wavelengths.  The objects were selected according to dust luminosity, with values at least $100\times$ above the stellar luminosity at either 60$\mu$m (IRAS) or 70$\mu$m (\textit{Spitzer}) for all sources in this sample \citep[e.g.][]{rhee07,carp09,che11,che12}. Along with the CO data that were the primary goal of the survey, we simultaneously collected sensitive continuum data at 0.5-1'' ($\sim$50\,AU) resolution, incidentally providing the largest uniform sample of spatially resolved debris disk observations at millimeter wavelengths to date (described in Section~\ref{sec:observations}).  

Here we present the results of our observations of the 24 disks in the sample (23 debris disks plus one additional source since determined to be a protoplanetary rather than a debris disk; see Section~\ref{sec:results}), along with an analysis of the ALMA continuum visibilities and the broad-band SED collected from the literature that constrains the basic spatial distribution and grain properties of the millimeter dust in the sample (Section~\ref{sec:analysis}).  We identify three strong detections of CO(2-1) emission from debris disk-hosting stars -- two new systems and one previously identified -- along with two additional tentative detections, but we defer detailed analysis of the CO observations to an accompanying publication \citep{hal16}.  We discuss the implications of the continuum analysis for the underlying properties of planetary systems and compare with surveys at optical and infrared wavelengths (Section~\ref{sec:discussion}), and then summarize the major results (Section~\ref{sec:conclusions}).  

\section{Observations}
\label{sec:observations}

We obtained ALMA Cycle 1 observations of 24 sources over a total of 6 nights between 2013 Dec 14 and 2014 Dec 14.  Table \ref{tab:obsparam} lists the date of observation, time on source for each target in the field, number of antennas, minimum and maximum projected baseline lengths, median precipitable water vapor (PWV; a measure of atmospheric transparency), the flux calibrator, the passband calibrator, and the gain calibrator for each night. The sample of disks was subdivided into 5 fields based on proximity, with one in Upper Sco, two in Lower Centaurus Crux, and two in Upper Centaurus Lupus (Table \ref{tab:obsResults}). The absolute flux scale, set by observations of a Solar system object with a high-quality flux model (either Ceres or Titan for this sample), is subject to an assumed $20\,\%$ systematic uncertainty due to the typical uncertainty in flux models of these solar system objects. The passband calibrator is a bright quasar observed at sufficient signal-to-noise to calibrate irregularities in the spectral response of the receiver.   The gain calibrator is a quasar located close on the sky to the sources of interest, observations of which are interleaved with observations of the target sources to calibrate variations in the atmospheric and instrumental amplitudes and phases of the interferometer response.  Band 6 was utilized for all observations, with four dual-polarization spectral windows centered at frequencies of 230.6, 232.6, 245.4, and 247.4\,GHz. The first spectral window, containing the $^{12}$CO J=2-1 line at a frequency of 230.53800\,GHz, has a bandwidth of 234.4\,MHz per polarization and a channel spacing of 122.1\,KHz (0.16\,km\,s$^{-1}$). The other three spectral windows, each with a bandwidth of 1.875\,GHz per polarization, were aggregated into a total 5.625\,GHz bandwidth (per polarization) for continuum analysis. 

CASA reduction scripts for the ALMA data were provided by the staff at the North American ALMA Science Center (NAASC).  The data reduction steps applied include phase correction with 183\,GHz water vapor radiometers, bandpass calibration, flux calibration, and gain calibration.  For flux calibration, we adopted the Butler-JPL-Horizon 2012 models in CASA version 4.1 or 4.2.  For the spectral line images, every four channels in the visibility data were averaged to provide channels with 244\,kHz resolution and sampling.  Before imaging the spectral line, the continuum emission was subtracted using the line-free regions of the spectrum.  
 
\begin{table*}
\caption{Observational Parameters}
\resizebox{\textwidth}{!}{
\begin{tabular}{lccccccccc}
\hline
\hline
\multicolumn{1}{c}{Field} & Date & Number of & Time on & Number of & Baseline & Median PWV & \multicolumn{3}{c}{\underline{$~~~~~~~~~~~~~~$Calibrators$~~~~~~~~~~~~~~$}}\\
&         & Sources & Source (min) & Antennas & Lengths (m) &  $\pm 1 \sigma$ (mm) & Flux & Passband & Gain \\
 \hline
1) Lower Centaurus Crux  & 2013 Dec 14 & 4 & 8.77 & 26 & 15-445 & $0.81$ & Ceres & J1107-4449 & J1424-4903 \\    
2) Upper Centaurus Lupus & 2014 Jan 10 & 5 & 10.2 & 26 & 15-290 & $2.68$ & Ceres & J1427-4206 & J1457-3539  \\   
3) Upper Centaurus Lupus & 2014 Jan 25 & 4 & 10.2 & 26 & 15-399 & $0.82$ & Titan & J1626-2951 & J1636-4102  \\     
4) Upper Scorpius & 2014 Mar 23-24 & 5 & 6.88 & 36 & 15-438 & $2.84$ & Titan  & J1517-2422 & J1626-2951 \\         
5) Lower Centaurus Crux & 2014 Dec 14 & 6 &  9.73 & 28 & 15-1284 & $0.69$ & Titan & J1427-4206 & J1112-5703 \\  
\hline
\label{tab:obsparam}
\end{tabular}
}
\end{table*}

\section{Results}
\label{sec:results}

\subsection{1240\,$\mu$m Continuum}

\begin{figure*}[ht!]
\centering
\includegraphics[height= .7\textheight]{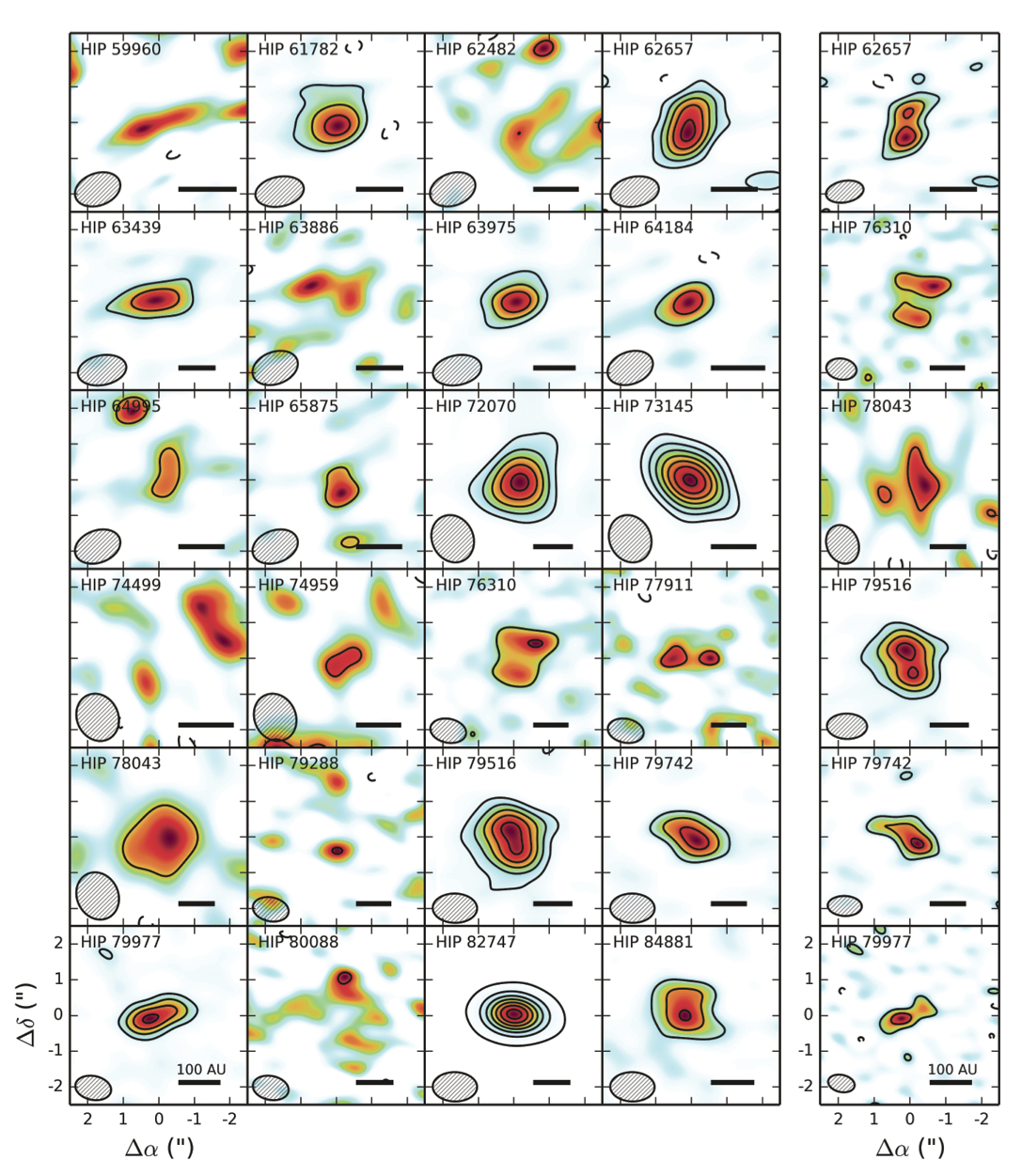}
\caption{Contour maps of continuum emission at 1240\,$\mu$m. Fluxes, beam characteristics, the RMS noise ($\sigma$) in each image, and the peak detection ($\times \sigma$) are given for all disks in Table \ref{tab:obsResults}. The rightmost column shows selected disks imaged with lower Briggs weighting parameters, placing more emphasis on long baselines to highlight possible inner cavities; corresponding imaging characteristics are also given in Table \ref{tab:obsResults}. The contours are [-3, 3, 6, 9, ...] $\times \sigma$ for all disks except HIP 82747, which has contours at [5, 25, 45, 65, 85, 105, 125] $\times \sigma$. Solid contours indicate positive flux densities, whereas dashed contours indicate negative flux densities. The color map is scaled from zero flux (white) to the peak flux (dark red) in each image.} 
\label{fig:johnFP}
\end{figure*}

A total of 20 of the 24 sources were detected at the $3\sigma$ level by our ALMA observations, according to the peak SNR in the naturally weighted images. The remaining four sources were marginally detected at the 2.7-2.8$\sigma$ level, although we do not conduct any further analysis on these sources.  Figure~\ref{fig:johnFP} displays naturally weighted images of the full sample at a wavelength of 1240\,$\mu$m, with separate panels using different weighting schemes to highlight structure in sources that seem to have resolved inner edges.  Table~\ref{tab:obsResults} provides details of the imaging parameters used to create these insets.

Table \ref{tab:obsResults} summarizes the basic continuum results, including the total measured flux density of the disk, naturally weighted beam size and position angle, RMS noise in a naturally weighted image, and peak SNR of each detection. Total fluxes were measured for unresolved sources by fitting a point source using the MIRIAD\footnote{Multichannel Image Reconstruction, Image Analysis and Display; see \citet{sau95} for more information.} task \texttt{uvfit}, whereas fluxes for resolved sources were estimated by fitting an elliptical Gaussian. The derived values are all consistent with expected values for debris disks at this wavelength except for HIP 82747 (AK Sco), which is in fact an optically thick, circumbinary protoplanetary disk \citep[see, e.g.,][]{jan15,cze15}. The centroid positions are consistent with the expected position of the star at the time of observation for all of the sources except HIP 72070, for which an offset of $\Delta \alpha = -0.13'', \Delta \delta = -0.08''$ is noted.  According to the absolute pointing accuracy quoted in the ALMA Cycle 4 Technical Handbook\footnote{https://almascience.nrao.edu/proposing/technical-handbook}, this is a 2$\sigma$ difference from the expected position, which is likely to occur spuriously in a sample of 20 objects.

The MIRIAD task \texttt{uvfit} was utilized to determine whether each disk was spatially resolved by our observations and to measure the total flux density of each detected source. We first conducted a fit of each disk with an elliptical Gaussian in the visibility domain; if the major axis length was measured with a signal-to-noise ratio (SNR) of $> 3\sigma$, we consider the disk spatially resolved and are able to estimate the position angle ($PA$) and place a lower limit on the inclination ($i$). Sources that were also resolved along the minor axis allow us to estimate both the position angle and inclination of the disk. The nine sources that were only resolved along the major axis are HIP~63439, HIP~61782, HIP~63975, HIP~62657, HIP~72070, HIP~78043, HIP~84881, HIP~79742, and HIP~79977. The four sources that were resolved along both the major 
and minor axis are HIP~73145, HIP~79516, HIP~82747, and HIP~76310.  The remaining
 seven sources are spatially unresolved.  
While we used the MIRIAD uvfit results to determine which disks are 
spatially resolved and therefore warrant further analysis, we do not 
report the PA and inclination values derived from the uvfit task, since 
they are superseded by the MCMC analysis described in Section~\ref{sec:analysis}.  PA 
and inclination values from the MCMC analysis are given in Tables~\ref{tab:unresolved} and \ref{tab:resolved}.

\begin{table*}[ht!]
\caption{Continuum Measurements and Imaging Parameters}
\resizebox{\textwidth}{!}{
\begin{tabular}{lcccccc}
\hline
\hline
\multicolumn{1}{c}{Source} & Field & $S_{total}$ ($\mu$Jy)& Beam Size ($''$) & Beam PA ($^{\circ}$)& $\sigma$ ($\mu$Jy\,bm$^{-1})$ & Peak S/N \\ 
 & Number & & \multicolumn{3}{c}{\underline{Naturally Weighted Images}} &  \\ 
\multicolumn{1}{c}{(\RNum{1})} & (\RNum{2}) & (\RNum{3}) & (\RNum{4}) & (\RNum{5}) & (\RNum{6}) & (\RNum{7}) \\
\hline
HIP 59960 & 5 & ... & $1.33 \times 0.87$ & $110$ & 44 & 2.8 \\    
HIP 61782 & 1 &  $710 \pm 110^{a}$ & $1.36 \times 0.83$ & $102$ & 41 & 10.6 \\     
HIP 62482 & 5 & $130 \pm 50$ & $1.30 \times 1.00$ & $104$ & 44 & 3.0 \\   
HIP 62657 & 1 & $1290 \pm 110^{a} $ & $1.37 \times 0.83$ & $101$ & 43 & 16.1 \\     
HIP 63439 & 1 & $520 \pm 90^{a}$ & $1.38 \times 0.83$ & $101$ & 44 & 8.6\\     
HIP 63886 & 5 & ... & $1.31 \times 0.88$ & $109$ & 57 & 2.7 \\     
HIP 63975 & 1 & $620 \pm 80^{a}$ & $1.39 \times 0.81$ & $99$ & 45 & 11.7\\     
HIP 64184 & 5 & $430 \pm 50$ & $1.32 \times 0.89$ & $110$ & 54 & 7.8 \\    
HIP 64995 & 5 & $180 \pm 50$ & $1.32 \times 0.88$ & $109$ & 45 & 4.0 \\   
HIP 65875 & 5 & $270 \pm 50$ & $1.32 \times 0.86$ & $108$ & 48 & 5.2 \\  
HIP 72070 & 2 & $1460 \pm 150^{a}$ & $1.39 \times 1.16$ & $28$ & 62 & 16.1 \\     
HIP 73145 & 2 & $2900 \pm 150^{a} $&  $1.36 \times 1.16$ & $28$ & 90 & 21.8 \\     
HIP 74499 & 2 & ... & $1.35 \times 1.16$ & $28$ & 66 & 2.8 \\     
HIP 74959 & 2 & $160 \pm 70$ & $1.36 \times 1.18$ & $24$ & 50 & 4.0 \\    
HIP 76310 & 4 & $1200 \pm 200^{a}$ & $0.97 \times 0.67$ & $81$ & 51 & 7.1 \\ 
HIP 77911 & 4 & $130 \pm 50$ & $1.00 \times 0.67$ & $79$ & 47 & 4.0 \\    
HIP 78043 & 2 & $340 \pm 70^a$ & $1.35 \times 1.22$ & $20$ & 69 & 5.1\\  
HIP 79288 & 4 & $200 \pm 60$ & $1.02 \times 0.67$ & $80$ & 61 & 3.2 \\    
HIP 79516 & 3 & $1850 \pm 120^{a}$ & $1.25 \times 0.82$ & $88$ & 45 &16.2 \\     
HIP 79742 & 3 & $880 \pm 90^{a}$ & $1.26 \times 0.78$ & $88$ & 51 & 10.6 \\     
HIP 79977 & 4 & $1300 \pm 120^{a}$ & $1.05 \times 0.67$ & $78$ & 58 & 12.7 \\   
HIP 80088 & 4 & ... & $1.06 \times 0.67$ & $78$ & 52 & 2.8 \\ 
HIP 82747 (AK Sco) & 3 & $35930 \pm 150^{a}$& $1.22 \times 0.76$ & $88$ & 240 &135 \\     
HIP 84881 & 3 & $720 \pm 110^{a}$ & $1.25 \times 0.82$ & $87$ & 40 & 9.4\\    

\hline
\hline
 & & & \multicolumn{3}{c}{\underline{Briggs Weighted Images}} &  \\ 
HIP 62657 & & & $1.06 \times 0.60^{b}$ & $101^{b}$ & $54^{b}$ \\
HIP 76310 & & &  $0.85 \times 0.59^{c}$ & $82^{c}$ & $55^{c}$ \\
HIP 78043 & & & $1.11 \times 0.88^{b}$ & $23^{b}$ & $65^{b}$ \\
HIP 79516 & & & $1.12 \times 0.71^{c}$ & $89^{c}$ & $49^{c}$ \\
HIP 79742 & & & $0.95 \times 0.56^{b}$ & $88^{b}$ & $47^{b}$ \\
HIP 79977 & & & $0.75 \times 0.48^{c}$ & $78^{c}$ & $77^{c}$ \\
\hline
\multicolumn{7}{p{\textwidth}}{$~~~~$\textsc{Note} \textemdash} \\
\multicolumn{7}{p{\textwidth}}{Column \RNum{1}: Source name.} \\
\multicolumn{7}{p{\textwidth}}{Column \RNum{2}: The field number as denoted by chronological order of observation (see Table \ref{tab:obsparam}).}\\
\multicolumn{7}{p{\textwidth}}{Column \RNum{3}: Integrated flux density measured by fitting a point source to the visibilities using the MIRIAD command \texttt{uvfit} (unless otherwise indicated), with ellipses denoting the source was not detected at the 3$\sigma$ level.}\\
\multicolumn{7}{p{\textwidth}}{Column \RNum{4}: FWHM beam size for the images in Figure \ref{fig:johnFP}.}\\
\multicolumn{7}{p{\textwidth}}{Column \RNum{5}: Beam position angle, measured east of north.}\\ 
\multicolumn{7}{p{\textwidth}}{Column \RNum{6}: RMS noise for the naturally weighted images, measured across many beam widths off the position of the disk.}\\
\multicolumn{7}{p{\textwidth}}{Column \RNum{7}: Peak SNR of the disk relative to the RMS noise level.}\\
\multicolumn{7}{p{\textwidth}}{$^a$ : Integrated flux density measured by fitting an elliptical Gaussian to the visibilities in the case of a resolved disk.} \\
\multicolumn{7}{p{\textwidth}}{$^b$ : Imaged with a Briggs weighting parameter of 0.} \\
\multicolumn{7}{p{\textwidth}}{$^c$ : Imaged with a Briggs weighting parameter of 0.5.}
\end{tabular}
}
\label{tab:obsResults}
\end{table*}

\subsection{CO(2-1)}
\label{sec:co21}

\begin{figure*}[h!]
\centering
\includegraphics[height= .7\textheight]{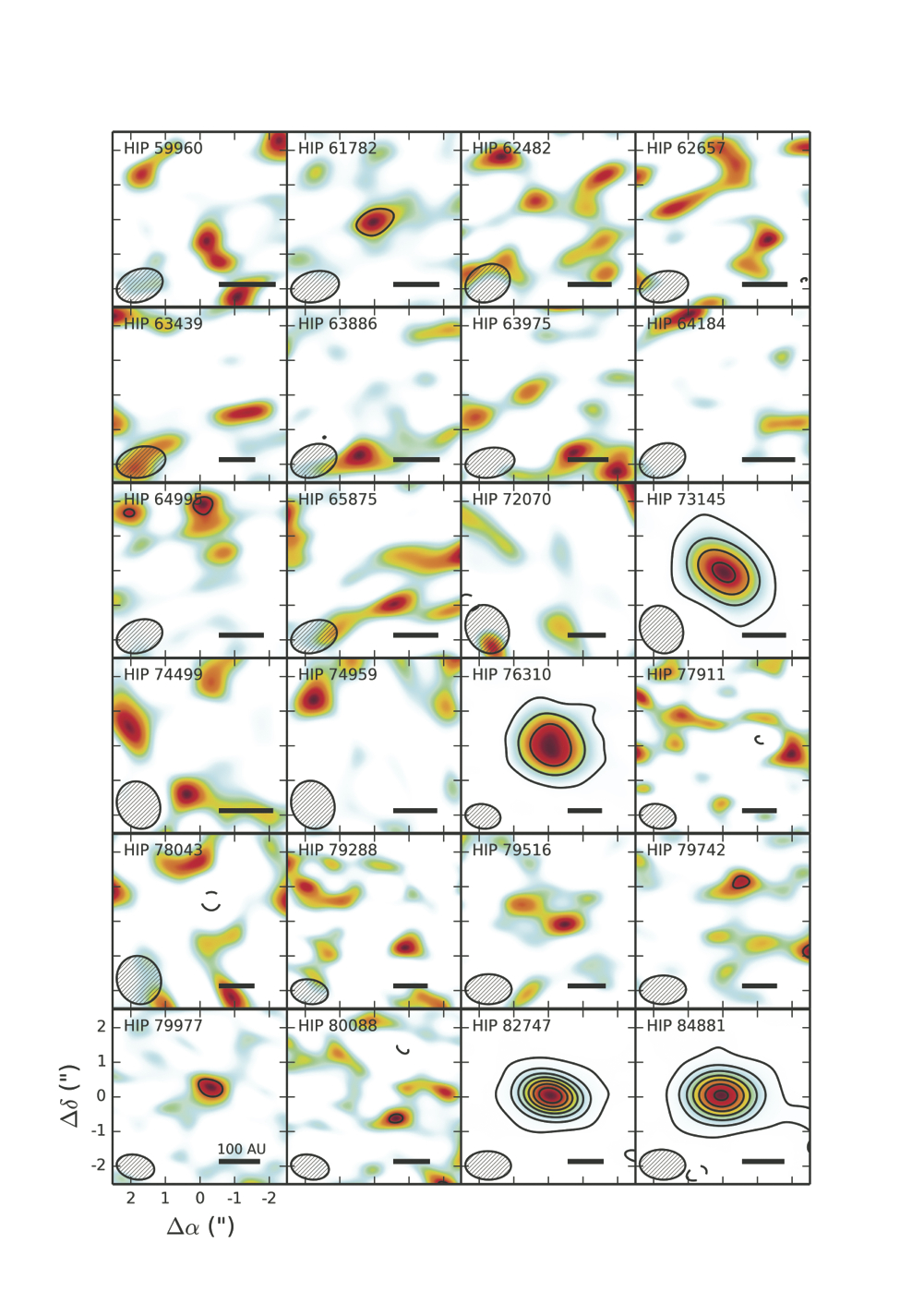}
\caption{Contour maps of the $^{12}$CO(2-1) integrated intensity.  Contours start at 3$\sigma$ with intervals of 10$\sigma$.  The RMS noise in the images and the velocity interval used to compute the integrated intensity are indicated in Table~\ref{tab:COresults}. The color map is scaled from zero flux (white) to the peak flux (dark red) in each image. } 
\label{fig:COintegrated}
\end{figure*}

\begin{figure*}[h!]
\centering
\includegraphics[width= .8\textwidth]{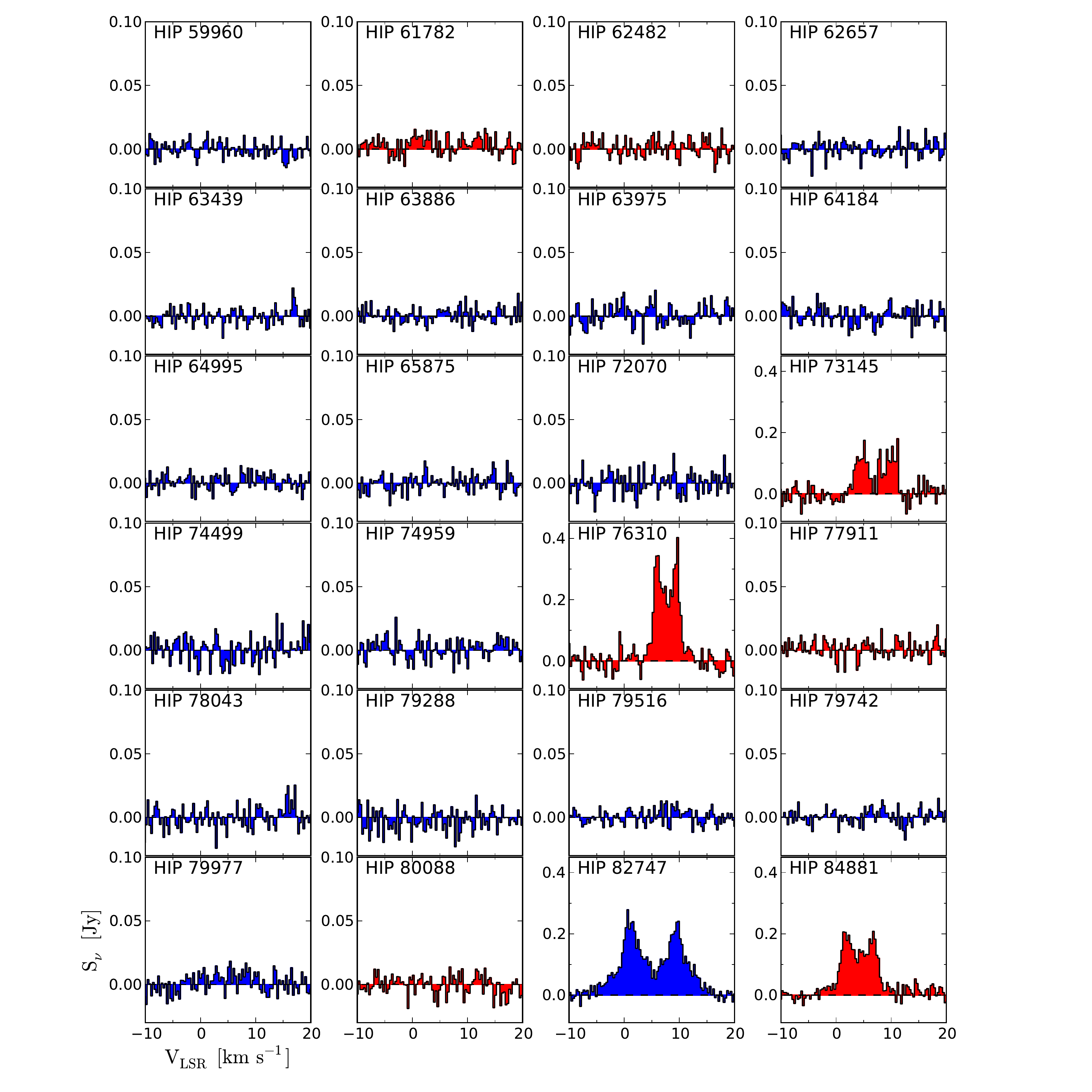}
\caption{$^{12}$CO(2-1) spectra toward the 24 sources in our sample. Spectra from B/A stars are displayed in red and spectra from F/G stars are shown in blue.  The spectra for the clear detections (HIP 76310, HIP 74881, and HIP 73145) were obtained with a 2$''$ radius aperture; a 0\farcs5 radius aperture was used for all other sources. HIP 61782 and HIP 79977 have tentative detections.
} 
\label{fig:COspectra}
\end{figure*}

Strong $^{12}$CO(2-1) emission is detected toward 4 of the 24 sources with SNR greater than 19 (HIP~76310, HIP~84881, HIP~73145, and HIP~82747).  Two of these (HIP~76310 and HIP~84881) are new CO detections around debris disk-hosting stars.  The third, HIP~73145, was recently discovered by \citet{moo15}, while the fourth -- HIP 82747, also known as AK Sco -- is a previously known CO-rich disk orbiting a pre-main sequence double-lined spectroscopic binary, making it more similar to a protoplanetary than a debris disk \citep[e.g.,][]{and89,cze15}.  Two other sources (HIP 61782 and HIP 79977) exhibit tentative 3$\sigma$ CO detections.  Table~\ref{tab:COresults} summarizes the measured $^{12}$CO(2-1) line intensities.  We also tabulate the signal-to-noise ratio of the detection, which is computed as the total flux divided by the uncertainty, where the uncertainty is derived from the rms noise multiplied by the square root of the number of independent pixels.  Figure~\ref{fig:COspectra} shows the $^{12}$CO(2-1) spectra of the 24 targets, and Figure~\ref{fig:COintegrated} shows the moment 0 (velocity-integrated intensity) maps.  

While the surface brightness properties of the CO emission will be examined and modeled in a forthcoming publication \citep{hal16}, a few immediately obvious trends are worth noting.  The three strong CO detections around debris disk-hosting stars represent three of the seven intermediate-mass (B- or A-type) stars in the sample, while only one of the 16 Solar-mass (F- or G-type) stars presents a tentative detection of CO emission (except for AK Sco, which is the only pre-main sequence star in the sample and is therefore not a good comparison object).  These detection statistics would appear to suggest that CO-rich debris disks are common around young intermediate-mass stars (occurring in $\sim$50\% of the small sample in this work) and rare around Solar-type stars.  It is unlikely that the difference arises exclusively from the higher temperatures in the gas disks induced by the presence of a hotter central star.  If the disks are in LTE, even a factor of two reduction in gas temperature at comparable radii would result in a factor of two lower flux in the Rayleigh-Jeans tail at millimeter wavelengths (for optically thick emission), which should be easily detectable at comparable CO masses given that the three strong CO detections all exhibit SNR $>19$.  While the flux is not always directly proportional to disk temperature (excitation, geometry, and optical depth also play a role), the large difference in flux between the disks detected at extremely high SNR and the non-detections suggests that the higher temperature of the A star disks is not the only reason for the higher detection fraction around those stars.  There is also no obvious trend between the spectral type of the central star and the total mass of dust in the disk that would predict a systematically lower mass of gas and dust in the debris disks around Solar-type stars.  

\begin{table*}[ht!]
\caption{CO J=2-1 Measurements and Imaging Parameters}
\resizebox{\textwidth}{!}{
\begin{tabular}{ccccccc}
\hline
\hline
\multicolumn{1}{c}{Source} & Beam Size ($''$) & $S_\mathrm{CO}$ & Beam PA ($^{\circ}$) & $\sigma_\mathrm{line}$ & $\sigma_\mathrm{int}$ &  S/N\\ 
 &  & (mJy km s$^{-1}$) & & (mJy beam$^{-1}$) & (mJy beam$^{-1}$ km s$^{-1}$) &  \\ 
(\RNum{1}) & (\RNum{2}) & (\RNum{3}) & (\RNum{4}) & (\RNum{5}) & (\RNum{6}) & (\RNum{7}) \\
\hline
HIP 59960  & $ 1.37\times 0.93$ & 109 &   6.4 &    15 &    13 $\pm$    16 &  0.8\\
{\bf HIP 61782}  & $ 1.41\times 0.88$ & 102 &   6.5 &    16 &    92 $\pm$    17 &  5.5\\
HIP 62482  & $ 1.34\times 1.05$ & 116 &   6.7 &    14 &     3 $\pm$    14 &  0.2\\
HIP 62657  & $ 1.42\times 0.88$ & 101 &   6.7 &    14 &     8 $\pm$    16 &  0.5\\
HIP 63439  & $ 1.43\times 0.89$ & 101 &   6.7 &    13 &    -0 $\pm$    14 & -0.0\\
HIP 63886  & $ 1.35\times 0.93$ & 109 &   6.7 &    13 &   -15 $\pm$    12 & -1.3\\
HIP 63975  & $ 1.43\times 0.86$ &  99 &   6.3 &    15 &     7 $\pm$    15 &  0.5\\
HIP 64184  & $ 1.36\times 0.95$ & 109 &   6.7 &    14 &     1 $\pm$    15 &  0.1\\
HIP 64995  & $ 1.36\times 0.93$ & 109 &   6.7 &    13 &    11 $\pm$    14 &  0.8\\
HIP 65875  & $ 1.36\times 0.92$ & 107 &   6.8 &    14 &     2 $\pm$    13 &  0.2\\
HIP 72070  & $ 1.45\times 1.20$ &  28 &   9.3 &    19 &     4 $\pm$    18 &  0.2\\
{\bf HIP 73145}  & $ 1.42\times 1.21$ &  28 &  10.1 &    16 &   798 $\pm$    35 & 22.5\\
HIP 74499  & $ 1.41\times 1.21$ &  28 &   8.9 &    16 &   -33 $\pm$    20 & -1.6\\
HIP 74959  & $ 1.42\times 1.22$ &  24 &   8.9 &    19 &    -9 $\pm$    18 & -0.5\\
{\bf HIP 76310}  & $ 1.02\times 0.71$ &  81 &   7.3 &    14 &  1406 $\pm$    78 & 18.0\\
HIP 77911  & $ 1.05\times 0.71$ &  79 &   6.8 &    14 &     8 $\pm$    18 &  0.4\\
HIP 78043  & $ 1.41\times 1.26$ &  20 &   8.9 &    17 &   -17 $\pm$    18 & -0.9\\
HIP 79288  & $ 1.07\times 0.71$ &  80 &   7.3 &    17 &   -25 $\pm$    19 & -1.3\\
HIP 79516  & $ 1.34\times 0.87$ &  90 &   5.6 &    11 &    15 $\pm$    11 &  1.3\\
HIP 79742  & $ 1.34\times 0.83$ &  91 &   5.7 &    12 &    14 $\pm$    12 &  1.2\\
{\bf HIP 79977}  & $ 1.10\times 0.70$ &  77 &   6.8 &    14 &    60 $\pm$    15 &  4.1\\
HIP 80088  & $ 1.11\times 0.71$ &  77 &   6.7 &    14 &     8 $\pm$    14 &  0.6\\
{\bf HIP 82747}  & $ 1.32\times 0.82$ &  88 &   5.8 &    23 &  2189 $\pm$   208 & 10.5\\
{\bf HIP 84881}  & $ 1.32\times 0.87$ &  87 &   5.9 &    11 &  1183 $\pm$    37 & 32.3\\
\hline
\multicolumn{7}{p{\textwidth}}{$~~~~$\textsc{Note} \textemdash Sources with CO detections (strong or marginal significance) are highlighted in boldface.} \\
\multicolumn{7}{p{\textwidth}}{Column \RNum{1}: Source name.} \\
\multicolumn{7}{p{\textwidth}}{Column \RNum{2}: Integrated CO J=2-1 intensity measured in the ALMA images. An aperture radius of 2$''$ was used for HIP~73145, HIP~76310, and HIP~84881, and a 0\farcs5 radius aperture for the remaining sources.} \\
\multicolumn{7}{p{\textwidth}}{Column \RNum{3}: FWHM beam size for the images in Figure \ref{fig:COintegrated}.} \\
\multicolumn{7}{p{\textwidth}}{Column \RNum{4}: Beam position angle, measured east of north.} \\
\multicolumn{7}{p{\textwidth}}{Column \RNum{5}: RMS noise in the CO J=2-1 spectral images per 0.32\,km\,s$^{-1}$ channel.} \\
\multicolumn{7}{p{\textwidth}}{Column \RNum{6}: RMS noise in the CO J=2-1 integrated intensity images measured in an annulus between 4 an 8\arcsec\ centered on the stellar position. For HIP~73145, HIP~76310, HIP~84881 and HIP~82747 the CO was integrated between 3 to 12, 5 to 10.6, -9 to 15, and 0.5 to 8.2\,km\,s$^{-1}$, respectively. For the two marginal detections (HIP~61782 and HIP~79977) the profiles were integrated between 0 and 10\,km\,s$^{-1}$ and 0 and 15\,km\,s$^{-1}$ respectively in order to maximize the signal-to-noise. All other sources were integrated between $\pm$5\,km\,s$^{-1}$ from their systemic velocities as listed in \citet[][]{kha07}.  Only for HIP~62482 was the systemic velocity unknown, so its spectrum was integrated between $-5$\,km\,s$^{-1}$ and $5$\,km\,s$^{-1}$.
}\\
\multicolumn{7}{p{\textwidth}}{Column \RNum{7}: Signal to noise ratio of the measured CO integrated intensity.}
\end{tabular}
}
\label{tab:COresults}
\end{table*}

\section{Analysis}
\label{sec:analysis}

For the 12 optically thin dust disks that are spatially resolved by our observations, we model the 1240\,$\mu$m ALMA visibilities and unresolved SED in order to constrain basic geometric properties of the disk and determine the characteristics of the constituent dust grains.We follow the modeling method and assumptions described in \citet{ric13}, the essentials of which we briefly recap below.   We exclude AK Sco from this sample because its disk is optically thick and violates the assumption of low optical depth in our model.  

The combination of spatially resolved visibilities and SED breaks the degeneracy between the distance of a dust grain from the star and its temperature, which is determined by its size (in the context of assumed optical properties). By combining temperature information from the SED with position information from the visibilities, we can learn about disk structure and the basic properties of the grain size distribution. For the resolved sources, we describe the surface density of each disk with a single power law $p$ such that $\Sigma(r) \propto r^{-p}$ with abrupt cutoffs at the inner radius ($R_\mathrm{In}$) and outer radius ($R_\mathrm{Out}$, modeled as $R_\mathrm{In} + \Delta R$). However, because the beam size is large compared to the typical belt widths, we are unable to break the degeneracy between $R_\mathrm{Out}$ and $p$ that arises in our modeling \citep[this is a well-known degeneracy; see discussion in][]{ric13}. As such, we set $p = 1$, a typical surface density profile observed in protoplanetary disks \citep{andr09}, and fit only the outer radius.  There are very few observations constraining surface brightness profiles in debris disks, although in some cases there is evidence that the surface brightness might increase with radius \citep[see, e.g.][]{mac13,ric15}; however, since the disk width is unresolved for all but a handful of sources, we do not expect this choice to significantly affect the derived disk properties.  Our optically-thin models have a mass $M_\mathrm{Disk}$, inclination $i$, and position angle $PA$. 

\subsection{Disk Geometry from Visibility-Only Fits}

Because the disks are detected in the continuum with a SNR of only $\sim3-22$, we isolate as much spatial information as possible in a visibility-only fit before simultaneously modeling the relatively high-quality photometry and low-quality visibilities to probe grain characteristics. We vary the geometric parameters of the dust grains that give rise to the submillimeter emission in the outer disk ($R_\mathrm{In}$, $\Delta R$, $i$, $PA$) for these visibility-only models before fitting both the SED and visibilities. Otherwise, the SED dominates the fit and implies that we have better constraints on $R_\mathrm{In}$ and $\Delta R$ than we really do.  The parameters $R_\mathrm{In}$ and $\Delta R$ are fundamentally spatial parameters, but they can be influenced by the need to recreate a specific range of grain temperatures to reproduce the shape of the SED; this may instead be carried out by adding a second population of smaller grains that have a different temperature but do not emit efficiently enough at long wavelengths to contribute flux to the ALMA image.  We discuss this possibility further in Section~\ref{sec:SEDfit} below. 

We create high-resolution model images of geometrically and optically thin disks at the $1240\,\mu$m wavelength of the ALMA observations, including a variable disk mass $M_\mathrm{Disk}$ to scale the flux to match the ALMA data (assumptions about the grain opacities are described in Section~\ref{sec:SEDfit} below).  We set the resolution of the images to be approximately 1$\%$ the spatial scale sampled by the longest baselines in the data, i.e., 1AU/pixel, and sample these synthetic images at the same baseline separations and orientations as the data using the MIRIAD task \texttt{uvmodel}.  We then compare our model disks to the data in the visibility domain and calculate a $\chi^2$ metric of the goodness of fit.  We carry out this analysis in the visibility domain because the noise is well understood \citep{schw78}, whereas the images created using the non-linear \texttt{CLEAN} algorithm do not have well characterized uncertainties.  In addition, fitting to the visibilities preserves information from the longest baselines (corresponding to the smallest angular scales), whereas the resolution of the \texttt{CLEAN} image is always coarser than the smallest angular scale. 

In order to explore the uncertainties associated with each parameter, we utilize the affine-invariant Markov Chain Monte Carlo (MCMC) fitting technique as described by \cite{good10} and implemented in Python as \texttt{emcee} by \cite{fore13}. Using an MCMC method allows us to probabilistically sample the full parameter space described by our models, obtain a best-fit result, and get statistically robust error bars by characterizing the full posterior distribution function for each parameter. 

For the gas-rich disks HIP 73145, HIP 76310, and HIP 84881 \citep{hal16}, we set best fit values of $i$ and $PA$ from fits to the CO J=2-1 line emission before varying the rest of the disk geometry.  Since the SNR of the CO data is higher (except for HIP 73145) and the Keplerian rotation is resolved in the spectral domain, the constraints on disk geometry from CO are stronger than those from the continuum.  

$R_\mathrm{In}$ and $\Delta R$ are unresolved for five sources (HIP 63439, HIP 61782, HIP 63975, HIP 72070, and HIP 78043), but we note a strong degeneracy between these parameters in our MCMC modeling and that the best fit values of $R_\mathrm{In}$ and $\Delta R$ is larger than the beam in each case. Indeed, we find that the outer radius ($R_\mathrm{Out} = R_\mathrm{In} + \Delta R$) is resolved by the observations. $i$ and $PA$ are not well constrained for these sources. HIP 84881 has a marginally resolved inner radius and resolved width. We report $R_\mathrm{In}$, $\Delta R$, $R_\mathrm{Out}$, $i$, and $PA$ for these disks in Table \ref{tab:unresolved}.

\begin{table*}[ht!]
\caption{Model Parameters for Disks with Unresolved Inner Radii} 
\resizebox{\textwidth}{!}{
\def\arraystretch{1.2}
\begin{tabular}{ccccccccccccccccc}
\hline \hline
 & \multicolumn{10}{c}{\underline{$~~~~~~~~~~~~~~~~~~~~~~~~~~~~~~~~~~~~~~~~~~~~~$Visibility Only Fit $~~~~~~~~~~~~~~~~~~~~~~~~~~~~~~~~~~~~~~~~~~~~~$}}& \multicolumn{6}{c}{\underline{$~~~~~~~~~~~~~~~~~~$Simultaneous Visibility \& SED Fit$~~~~~~~~~~~~~~~~~~$}} \\ 
     & \multicolumn{2}{c}{\underline{$~~~~~~~~~R_\mathrm{In}$ [AU]$~~~~~~~~~$}} & \multicolumn{2}{c}{\underline{$~~~\Delta R$ [AU]$~~~$}} & \multicolumn{2}{c}{\underline{~~~$R_\mathrm{Out}$ [AU]$~~~$}} & \multicolumn{2}{c}{\underline{$~~~~~i$ [$^\circ$]$~~~~~$}} & \multicolumn{2}{c}{\underline{$~~~~~~PA$ [$^\circ$]$~~~~~~$}} & \multicolumn{2}{c}{\underline{$~~~R_\mathrm{In}$ [AU]$~~~$}} & \multicolumn{2}{c}{\underline{$~~~$log($M_\mathrm{Disk}$ [$M_\earth$])$~~~$}} & \multicolumn{2}{c}{\underline{$~~~$log($a$ [$\mu$m])$~~~$}} \\ 
    Source & Median $\pm\,\sigma$ & Best Fit & M$\pm\,\sigma$ & BF & M$\pm\,\sigma$ & BF & M$\pm\,\sigma$ & BF & M$\pm\,\sigma$ & BF & M$\pm\,\sigma$ & BF & M$\pm\,\sigma$ & BF & M$\pm\,\sigma$ & BF  \\

     HIP 61782 &      $ < 30 $ & 10 &     $ < 80 $ & 110 &      80$^{+40}_{-20}$ & 120 &     $ Unconst.$ & 50 &      $ Unconst. $ & $-30$ &     2.1$^{+0.3}_{-0.2}$ & 2.1 &     $-2.56^{+0.07}_{-0.08}$ & $-2.54$ &     0.67$^{+0.11}_{-0.09}$ & 0.70 \\

    HIP 63439 &    $ < 30 $ & 70 &    $ < 150 $ & 50 &    140$^{+90}_{-70}$ & 120 &    $ > 67 $ & 83 &    96$^{+15}_{-17}$ & 94 &    6.9$^{+1.5}_{-0.9}$ & 6.8 &    $-2.01^{+0.10}_{-0.10}$ & $-1.98$ &    1.4$^{+0.2}_{-0.2}$ & 1.4 \\

     HIP 63975 &     $ < 14 $ & 2 &    $ < 80 $ & 120 &     70$^{+30}_{-30}$ & 120 &    $ Unconst. $ & 70 &     $ Unconst. $ & $-30$ &     0.36$^{+0.02}_{-0.02}$ & 0.37 &     $-2.80^{+0.04}_{-0.02}$ & $-2.80$ &     $-0.59^{+0.15}_{-0.14}$ & $-0.64$ \\

     HIP 72070 &     $ < 40 $ & 10 &     $ < 110 $ & 150 &      110$^{+50}_{-30}$ & 160 &     $ > 50 $ & 70 &     $-59^{+17}_{-12}$ & $-62$ &     5.7$^{+1.2}_{-0.8}$ & 5.4 &     $-1.75^{+0.05}_{-0.05}$ & $-1.73$ &     0.51$^{+0.07}_{-0.08}$ & 0.54 \\
        
    HIP 78043 &    $ < 70 $ & 50 &    $ < 190 $ & 110 &180$^{+170}_{-60}$ & 160 &    $ Unconst. $ & 20 &    $ Unconst. $ & $-50$ &    $6.3^{+0.8}_{-0.6}$ & 6.2 &    $-1.97^{+0.14}_{-0.13}$ & $-1.89$ &    $1.3^{+0.3}_{-0.2}$ & 1.4 \\
        
    HIP 84881 &     $ < 20 $ & 10 &     130$^{+40}_{-30}$ & 130 &    150$^{+30}_{-30}$ & 140 &   --$^a$  & 30 &   --$^a$  & $-86$ &     2.96$^{+0.07}_{-0.11}$ & 2.99 &     $-2.48^{+0.04}_{-0.03}$ & $-2.49$ &     $-0.55^{+0.16}_{-0.15}$ & $-0.54$ \\
\hline
\multicolumn{17}{p{\textwidth}}{$~~~~$\textsc{Note} \textemdash $^a$ Because the CO line emission provides more stringent constraints on position angle and inclination, we fixed $PA$ and $i$ for these disks to the best-fit values reported in \citet{hal16}.}  \\ 
\end{tabular}
}
\label{tab:unresolved}
\end{table*}

The visibility-only fits resolve $R_\mathrm{In}$, and constrain $i$, and $PA$ for six sources: HIP 62657, HIP 73145, HIP 79516, HIP 76310, HIP 79742, and HIP 79977.  $\Delta R$, however, is either marginally resolved (HIP 73145, HIP 79516, HIP 76310) or unresolved (HIP 62657, HIP 79742, HIP 79977) in these fits. $R_\mathrm{In}$, $i$, $PA$, and constraints on $\Delta R$ are reported in Table \ref{tab:resolved}.  For the disks that have been previously resolved in scattered light at higher angular resolution (HIP 61782, HIP 62657, HIP 73145, and HIP 79977), the values of $PA$ and $i$ that we derive are consistent with the previously determined geometry to within the uncertainties: scattered light observations yield a nearly edge-on debris disk at a $PA$ of 155$^\circ$ for HIP 61782 \citep{kas15}, an edge-on disk at a $PA$ of $\sim165^\circ$ for HIP 62657 \citep{dra16}, a $PA$ of 61.4$^\circ \pm$0.4$^\circ$ and $i$ of 75.1$^\circ$$^{+0.8^\circ}_{-0.9^\circ}$ for HIP 73145 \citep{hun15}, and a $PA$ of 114$^\circ\pm$0.3$^\circ$ and $i$ of 84$^\circ$$^{+2^\circ}_{-3^\circ}$ for HIP 79977 \citep{tha13}.

\begin{table*}[ht!]
\caption{Model Parameters for Disks with Resolved Inner Radii} 
\resizebox{\textwidth}{!}{
\def\arraystretch{1.2}
\begin{tabular}{ccccccccccccccccc}
    \hline \hline
 & \multicolumn{8}{c}{\underline{$~~~~~~~~~~~~~~~~~~~~~~~~~~~~~~~~$Visibility Only Fit$~~~~~~~~~~~~~~~~~~~~~~~~~~~~~~~~$}}& \multicolumn{8}{c}{\underline{$~~~~~~~~~~~~~~~~~~~~~~~~~~~~~~~~~~~~~~~$Simultaneous Visibility \& SED Fit$~~~~~~~~~~~~~~~~~~~~~~~~~~~~~~~~~~~~~~~$}} \\      
     & \multicolumn{2}{c}{\underline{$~~~~~~~~~R_\mathrm{In}$ [AU]$~~~~~~~~~$}} & \multicolumn{2}{c}{\underline{$~~~\Delta R$ [AU]$~~~$}} & \multicolumn{2}{c}{\underline{$~~~~~i$ [$^\circ$]$~~~~~$}} & \multicolumn{2}{c}{\underline{$~~~~PA$ [$^\circ$]$~~~~$}} & \multicolumn{2}{c}{\underline{$~~~R_{In,~Inner~Belt}$ [AU]$~~~$}} & \multicolumn{2}{c}{\underline{$~~~$log($M_\mathrm{Disk}$ [$M_\earth$])$~~~$}} & \multicolumn{2}{c}{\underline{$~~~$log($M_{Belt}$ [$M_\earth$])$~~~$}} & \multicolumn{2}{c}{\underline{$~~~$log($a$ [$\mu$m])$~~~$}} \\      
    Source & Median $\pm\,\sigma$ & Best Fit & M$\pm\,\sigma$ & BF & M$\pm\,\sigma$ & BF & M$\pm\,\sigma$ & BF & M$\pm\,\sigma$ & BF & M$\pm\,\sigma$ & BF & M$\pm\,\sigma$ & BF & M$\pm\,\sigma$ & BF \\
    HIP 62657 &     $45^{+15}_{-15}$ & 56 &     $< 90 $ & 60 & $ > 83 $ & 88  &     $-15^{+5}_{-5}$ &$ -14 $ &         1.71$^{+0.22}_{-0.20}$ & 1.70 &     $-1.93^{+0.04}_{-0.05}$ & $-1.94$ &    $-4.17^{+0.11}_{-0.11}$ & $-4.19$ &    0.52$^{+0.06}_{-0.06}$ & 0.51 \\

    HIP 73145 &      24$^{+11}_{-11}$ & 21 &      140$^{+30}_{-30}$ & 140 &   --$^a$  & 73 &  --$^a$    & 58 &     1.29$^{+0.12}_{-0.10}$ & 1.26 &    $-1.49^{+0.03}_{-0.03}$ & $-1.49$ &    $-3.45^{+0.08}_{-0.09}$ & $-3.47$ &    0.48$^{+0.04}_{-0.04}$ & 0.48 \\
    
     HIP 76310 &
    67$^{+20}_{-19}$ & 70 &
    $80^{+60}_{-50}$  & 70 & --$^a$
    & 28 & --$^a$
    & 48 &
    4.8$^{+0.5}_{-0.6}$ & 4.9 &
    $-2.08^{+0.07}_{-0.06}$ & $-2.12$ &
    $-3.68^{+0.11}_{-0.10}$ & $-3.77$ &
    $-0.4^{+0.2}_{-0.2}$ & $-0.5$ \\    

     HIP 79516 &      56$^{+11}_{-9}$ & 53 &     70$^{+20}_{-30}$ & 70 &     50$^{+6}_{-7}$ & 53 &     20$^{+7}_{-6}$ & 22 &     5.4$^{+1.1}_{-0.9}$ & 5.1 &     -1.67$^{+0.04}_{-0.04}$ & $-1.68$ &     -3.31$^{+0.16}_{-0.15}$ & $-3.40$ &     0.43$^{+0.06}_{-0.07}$ & 0.40 \\
  
    HIP 79742 & 
    73$^{+14}_{-19}$ & 83 &
    $ < 50 $ & 20 &
    $ > 72 $ & 79 &
     52$^{+5}_{-6}$ & 53 &
     5.4$^{+1.5}_{-1.1}$ & 5.9 &
     $-1.91^{+0.11}_{-0.06}$ & $-1.92$ &
     $-3.6^{+0.3}_{-0.3}$ & $-3.7$ &
     0.3$^{+0.2}_{-0.2}$ & 0.3 \\

    HIP 79977 & 
    60$^{+11}_{-13}$ & 71&
    $ < 50 $ & 20 &
    $ > 84 $ & 89 &
    $-65^{+3}_{-3}$ & $-66$ & 
    5.4$^{+1.2}_{-1.4}$ & 5.4 &
    $-2.00^{+0.09}_{-0.09}$ & $-1.99$ &
    $-3.62^{+0.15}_{-0.14}$ & $-3.62$ &
    0.0$^{+0.2}_{-0.2}$ & 0.0 \\
\hline   
\multicolumn{17}{p{\textwidth}}{$~~~~$\textsc{Note} \textemdash $^a$ Because the CO line emission provides more stringent constraints on position angle and inclination, we fixed $PA$ and $i$ for these disks to the best-fit values reported in \citet{hal16}.}  \\ 
\end{tabular}
}
\label{tab:resolved}
\end{table*}

\subsection{Simultaneous Modeling of the SED and Visibilities}
\label{sec:SEDfit}

Using the results from the visibility-only fits, we fix the geometry of the grain population responsible for the continuum emission from the outer disk and then perform simultaneous modeling of the ALMA visibilities and broad-band SED to place constraints on the basic properties of the dust grains (fixed and varied parameters are specified in Sections~\ref{sec:unresolved} and \ref{sec:resolved} below).  This analysis implicitly assumes that the small IR-emitting grains are spatially co-located with the larger grains that dominate emission in the submillimeter.  We fit models to the SED for data points with $\lambda$ $> 5\,\mu$m collected from the literature, as emission is dominated by the stellar photosphere at $\lambda$ $\lesssim 5\,\mu$m in debris disks. Mid-IR photometry was obtained with the IRS ($5.2 - 37.9\,\mu$m: systematic uncertainty assumed to be $10\%$), IRAC ($7.74\,\mu m$: $2\%$ systematic uncertainty), and MIPS ($24\,\mu m$: $2\%; 70\,\mu m$: $4\%$) instruments on \textit{Spitzer} \citep{carp06,carp09}, WISE \citep[W3, $11.56\,\mu m$: $4.5\%$; W4, 22.08$\,\mu m$: $5.7\%$,][]{jarr11,cut13}, and AKARI ($8.61\,\mu m$: $7\%$; \citealt{yam10,ish10}). The AKARI systematic uncertainty is a combination of two $5\%$ systematics related to flat fielding the data and issues with short exposures in the near-IR combined with the $2\%$ calibration uncertainty intrinsic to the instrument. The IRS spectrum originally consisted of $\sim$ 360 points, but was averaged down to 9 for the sake of computational efficiency. We do not include the ALMA flux in the SED fit to avoid lending it inappropriate weight in the fitting process, since the flux is implicitly included in the visibility $\chi^2$ calculation.  Absolute uncertainties were added with these systematics in quadrature for each measurement to generate appropriate uncertainties for modeling. 

We model the SED with two or three components: (1) a Kurucz-Lejeune model photosphere with solar metallicity Z = 0.01, (2) an extended, spatially-resolved debris disk, and when necessary, (3) an unresolved inner belt. Table~\ref{tab:blowoutGrains} lists the assumed stellar parameters (and corresponding references) for the Kurucz-Lejeune model photosphere and the calculated blowout grain size for each star in the sample. 

\begin{table*}[ht!]
\caption{Stellar Parameters}
\resizebox{\textwidth}{!}{
\begin{tabular}{ccccccc}
    \hline
    \hline
    Source & Spectral Type & $T_{Eff}$ (K) & Mass ($M_{\odot}$) & Luminosity ($L_{\odot}$) & Distance (pc) &  Blowout Size ($\mu$m) \\
    (\RNum{1}) & (\RNum{2}) & (\RNum{3}) & (\RNum{4}) & (\RNum{5})  & (\RNum{6}) & (\RNum{7}) \\
    \hline
    HIP 59960 & F5V & 6548 & 1.5$^{b}$ & 5.4 & 92 & 0.76 \\
    HIP 61782 & A0Vs & 8138 & 2.9 & 6.1 & 107 & 0.44\\
    HIP 62482 & A3III/IV & 8073 & 2.4 & 11.1 & 123 & 0.96 \\
    HIP 62657 & F5/6V & 6417 & 1.3$^{b}$ & 2.7 & 109 & 0.44 \\
    HIP 63439 & F3/5IV/V & 6617 & 1.4$^{b}$ & 3.7 & 143 & 0.56 \\
    HIP 63886 & F2V & 6871 & 1.5$^{b}$ & 5.0 & 107 & 0.70 \\
    HIP 63975 & F3/5V & 6955 & 1.4$^{c}$ & 4.4$^{c}$ & 123 & 0.66\\
    HIP 64184 & F3V & 6697& 1.5$^{b}$ & 3.2 & 85 & 0.45 \\
    HIP 64995 & F2IV/V & 6867 & 1.5$^{b}$ & 5.0 & 110 & 0.70 \\
    HIP 65875 & F6V & 6400 & 1.6$^{b}$ & 6.0 & 110 & 0.79 \\ 
    HIP 72070 & G1V & 5918 & 1.3$^{b}$ & 2.9 & 133 &0.47 \\
    HIP 73145 & A2IV & 8281 & 2.5 & 8.8 & 123 & 0.74 \\
    HIP 74499 & F3/5V & 6545 & 1.5$^{b}$ & 2.1 & 90 & 0.30\\ 
    HIP 74959 & F5V & 6374 & 1.3$^{b}$ & 2.7 & 133 & 0.44 \\
    HIP 76310 & A0V & 8883 & 2.9 & 23.1 & 151 & 1.7 \\
    HIP 77911 & B9V & 8685 & 3.4 & 32.8 & 148 & 2.0 \\
    HIP 78043 & F3V & 6639 & 1.5$^{b}$ & 4.3 & 144 & 0.61\\ 
    HIP 79516 & F5V & 6495 & 1.4$^{b}$ & 4.0 & 134 & 0.60 \\
    HIP 79288 & F0V & 6644 & 1.6$^{b}$ & 6.0 & 150 & 0.79 \\
    HIP 79742 & F6V & 6516 & 1.4$^{b}$ & 3.8 & 146 & 0.57 \\
    HIP 79977 & F2/3V & 6271 & 2.1$^{b}$ & 3.1 & 123 & 0.31 \\
    HIP 80088 & A9V & 6400 & 1.7 & 4.1 & 139 & 0.51 \\
    HIP 82747 & F5V & 6370$^{a}$ & 1.4 & 4.0 & 103 & 0.60 \\
    HIP 84881 & A0V & 8638 & 2.9 & 15.0 & 118 & 1.1 \\
    \hline
    \multicolumn{6}{p{\textwidth}}{$~~~~$\textsc{Note} \textemdash} \\
    \multicolumn{6}{p{\textwidth}}{Column \RNum{3}: Effective temperatures given by \cite{mcdo12}, unless noted.  Uncertainties are not specified by \citeauthor{mcdo12} but are at least $\sim 5\%$ due to a combination of the assumption of solar metallicity and unknown interstellar reddening for each source. In choosing a stellar photosphere model, we round these values to the closest multiple of 250, as this is the frequency of values given by \cite{leje97}.}\\
    \multicolumn{6}{p{\textwidth}}{Column \RNum{4}: The masses estimated from interpolating between the values given in \cite{alle00} from spectral type unless otherwise noted.}  \\
    \multicolumn{6}{p{\textwidth}}{Column \RNum{5}: Luminosities of each star as given by \cite{mcdo12}.}\\
    \multicolumn{6}{p{\textwidth}}{Column \RNum{6}: Distance to the star from {\it Hipparcos} parallax measurement \citep{lee07}}\\
    \multicolumn{6}{p{\textwidth}}{Column \RNum{7}: Blowout grain size as given by $a = \frac{3\,L_{\star}}{16 \pi \,\rho\,\text{G}\,M_{\star}\,c}$ for use in visibilities-only fits.} \\
    \multicolumn{6}{p{\textwidth}}{$^{a}$ \cite{czek15}.} \\
    \multicolumn{6}{p{\textwidth}}{$^{b}$ Masses as given by \cite{chen11} from isochrone fitting.} \\
    \multicolumn{6}{p{\textwidth}}{$^{c}$ The mass and luminosity of HIP 63975A rather than the binary, as the disk only surrounds HIP 63975A \citep{liss08}.} \\
    \hline
\end{tabular}
}
\label{tab:blowoutGrains}
\end{table*}

We assume that the composition of the grains is compact astrosilicates \citep{drai03} with characteristic grain size $a$.  We use realistic astrosilicate opacities and albedos generated using Mie theory (see \citealt{bohr83}) to determine grain temperatures, following the approach described in \cite{ric13}. We assume $\rho = 2.7$\,g\,cm$^{-3}$, which strikes a balance between the cometary and terrestrial materials assumed to make up astrosilicate grains \citep{blum08}. Taking the absorption and emission efficiency as a function of grain size and wavelength into account, the energy balance (and corresponding temperature) is solved for grain sizes between $0.1\,\mu$m and $3000\,\mu$m at 50\,AU. These values are then scaled by $r^{-1/2}$ to calculate the temperature of the grains at different distances from the central star. 

We model the emissive properties of the grains using the characteristic grain size $a$ and a long-wavelength power law index of grain emission efficiency $\beta$.  The emission efficiency of a grain as a function of wavelength, $Q_{\lambda}$, is modeled as $Q_{\lambda} = 1 - e^{-(\frac{\lambda}{2\pi a})^{-\beta}}$, which has $Q_{\lambda} \approx ({\lambda}/{2\pi a})^{-\beta}$ in the limit of $\lambda$ $>>$ $2\pi a$ and $Q_{\lambda} \approx 1$ when $\lambda$ $<<$ $2\pi a$. Such a ``modified blackbody" approach is common in the literature; for low-resolution data comparable to our own, this approach produces similar results to those obtained using a full grain size distribution \citep[see, e.g.,][]{wil04,paw14}.  Because we do not have the necessary long wavelength photometry to constrain $\beta$ for any of the disks in our sample, we set $\beta = 0.8$, a typical value for debris disks modeled using a similar approach \citep{stee15}.  This hybrid approach of using tabulated astrosilicate opacities for the temperature calculation while using a parameterized approximation of a grain size distribution for the emission efficiencies is not entirely internally self-consistent.  Nevertheless, it allows us to approximate the characteristics of a grain size distribution with sufficient computational efficiency to allow for robust statistical characterization of the model parameters using the computationally intensive MCMC method.  Using the tabulated astrosilicate opacites for the emission efficiency would increase the run time for each model by a factor of $\sim$30 and make the MCMC calculation intractable.  

\subsubsection{Disks with Unresolved Inner Radii}
\label{sec:unresolved}

For the disks that have unresolved inner radii (HIP 63439, HIP 61782, HIP 63975, HIP 72070, and HIP 78043), we vary $R_\mathrm{In}$ and $M_\mathrm{Disk}$ to fit the mid-IR fluxes in the SED while holding $i$ and $PA$ at their best fit values, even in cases where they are not well constrained. Because we resolve the outer radius of the disk ($R_\mathrm{Out,~Best~fit} = R_\mathrm{In,~Best~fit} + \Delta R_\mathrm{Best~fit}$), we allow $\Delta R$ to vary under the constraint that $\Delta R = R_\mathrm{Out,~Best~Fit} - R_\mathrm{In}$, ensuring that the outer radius will always be at the value as resolved by the visibilities. For these models, we assume that the grain size $a$ is equivalent to the blowout size (column \RNum{7} in Table \ref{tab:blowoutGrains}).  Column \RNum{2} in Table \ref{tab:AIC} reports the raw $\chi^{2}$ for these models. However, we find that we need to also vary $a$ in order to successfully recreate the SED and visibilities (raw $\chi^{2}$ in Column \RNum{3}) as justified by the Akaike Information Criterion \citep[AIC,][]{akai74}. The AIC is a statistical test that allows us to compare goodness of fit for two models using the $\chi^2$ statistic with appropriate penalties for models with additional parameters.  The models in which the grain size is included as a free parameter rather than fixed at the blowout size represent a significantly better fit to the data at the level reported in Column \RNum{4}.  This result likely reflects the influence of the grain size on the shape of the mid-IR SED, since it determines the inflection point beyond which the emission efficiency of the grains decreases below a value of 1.

\begin{table*}[ht!]
\caption{Significance of Models with Additional Parameters}
\resizebox{\textwidth}{!}{
\def\arraystretch{1}
\begin{tabular}{cccccccc}
    \hline
    \hline
    \multicolumn{4}{c}{\underline{$~~~~~~~~$Disks with Unresolved Inner Radii$~~~~~~~~$}} & \multicolumn{4}{c}{\underline{$~~~~~~~~~~~$Disks with Resolved Inner Radii$~~~~~~~~~~~$}} \\ 
    Source & Raw $\chi^{2}_{A}$ & Raw $\chi^{2}_{B}$ & Significance & Source & Raw $\chi^{2}_{A}$ & Raw $\chi^{2}_{B}$ & Significance \\
    (\RNum{1}) & (\RNum{2}) & (\RNum{3}) & (\RNum{4}) & (\RNum{5}) & (\RNum{6}) & (\RNum{7}) & (\RNum{8}) \\
    \hline
    HIP 61782 & 36774 & 36760 & 3.6$\sigma$ & HIP 62657 & 36715 & 36498 & $> 10\sigma$ \\
    HIP 63439 & 35834 & 35795 & 6.1$\sigma$ & HIP 73145 & 40910 & 40554 & $> 10\sigma$ \\
    HIP 63975 & 37285 & 37276 & 2.9$\sigma$ & HIP 76310 & 65853 & 65292 & $> 10\sigma$ \\
    HIP 72070 & 40069 & 40025 & 6.5$\sigma$ & HIP 79516 & 42764 & 42626 & $> 10\sigma$ \\
    HIP 78043 & 39550 & 39516 & 5.7$\sigma$ & HIP 79742 & 42942 & 42695 & $> 10\sigma$ \\
    HIP 84881 & 46208 & 43046 & $> 10\sigma$ & HIP 79977 & 67335 & 66909 & $> 10\sigma$ \\
    \hline
    \multicolumn{8}{p{\textwidth}}{\textsc{Note} \textemdash} \\
    \multicolumn{8}{p{\textwidth}}{Columns \RNum{2} and \RNum{3} compare models in which only the inner radius is allowed to vary (Column \RNum{2}) with models in which both the inner radius and characteristic grain size are allowed to vary (Column \RNum{3}).  Columns \RNum{6} and \RNum{7} compare models without an inner belt (Column \RNum{6}) to models in which there is an inner belt in addition to the outer disk resolved by the ALMA observations (Column \RNum{7}).} \\
\end{tabular}
}
\label{tab:AIC}
\end{table*}

\begin{figure*}[ht!]
\centering
\includegraphics[height= .65\textheight]{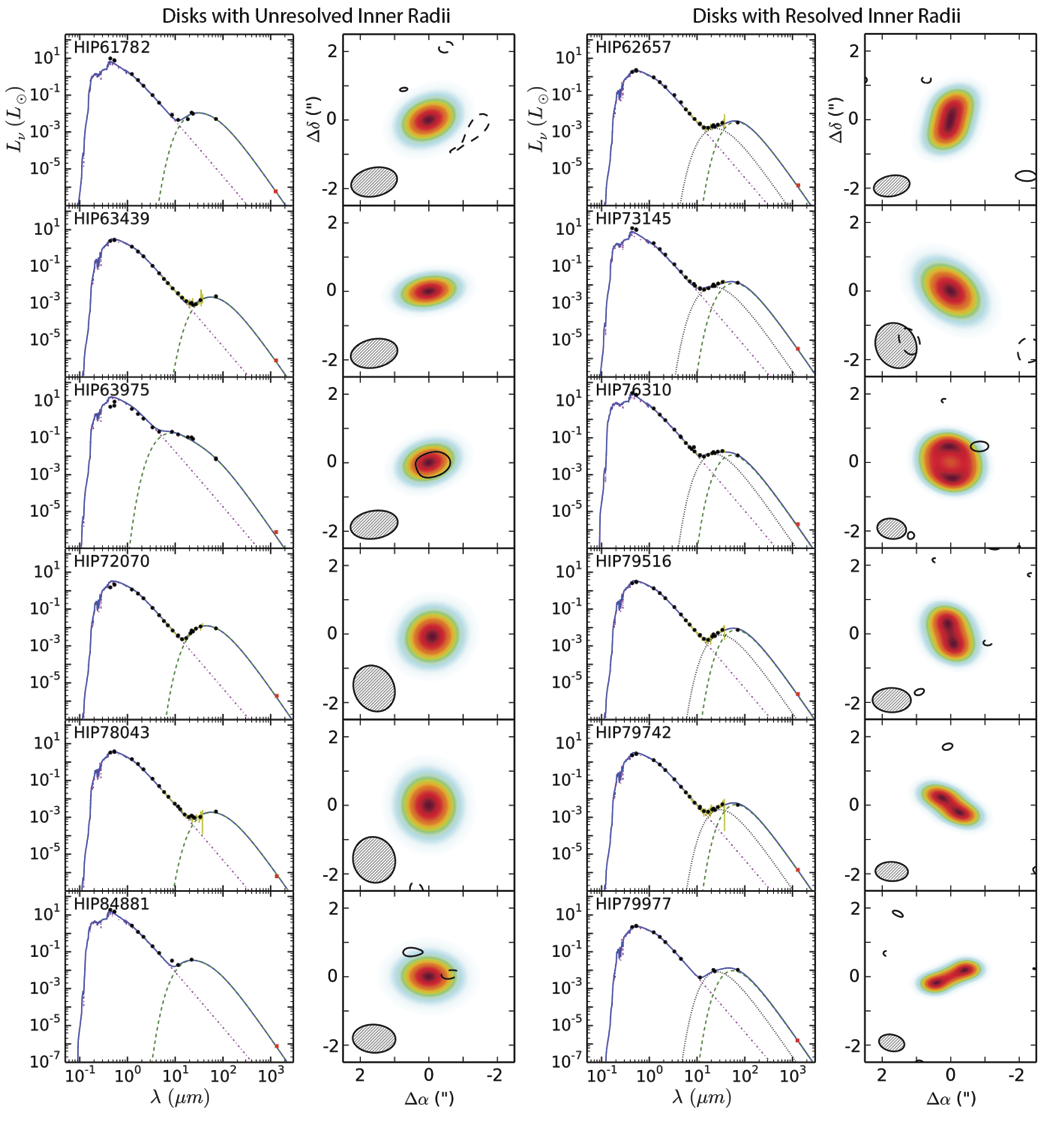}
\caption{Best fit SEDs, model images, and residual contours for all resolved disks. The SEDs display the best fit model (blue solid line), which is the sum of a Kurucz-Lejeune model photosphere (purple dotted line), disk model (green dashed line), and inner belt when necessary (gray dotted line). The model was fit to the observed photometry (black dots) for $\lambda > 5\,\mu$m, and the IRS spectrum (yellow) is shown for comparison purposes in disks for which these data have been collected.  The ALMA fluxes (red squares) are also shown for comparison, even though they were not included in the SED fit.  The color map displays the model images and is scaled from zero flux (white) to the peak flux (dark red). The contours show the residuals, with confidence levels of [-3, 3] $\times \sigma$. The model and residual images are naturally weighted for disks with unresolved inner radii and HIP 73145. Models and residual images for disks with resolved inner radii were imaged with differing weighting schemes except for HIP 73145 (see Table \ref{tab:obsResults}). The inner edge of HIP 73145's disk is only marginally resolved in the visibility domain by our models and is not visible in the images.}
\label{fig:johnSuperPlt}
\end{figure*}

For HIP 84881, which has a marginally resolved inner radius and resolved $\Delta R$ we fix the full geometry of the disk and attempt to model the SED by varying $a$ and $M_\mathrm{Disk}$ (raw $\chi^{2}$ in Column \RNum{2}), but find that varying only these parameters is insufficient to reproduce the observed data. We follow the same prescription as above, setting $\Delta R = R_\mathrm{Out} - R_\mathrm{In}$, and find that our models are successful if we allow $R_\mathrm{In}$, $a$ and $M_\mathrm{Disk}$ to vary (raw $\chi^{2}$ in Column \RNum{3}). Best fit and Median values $\pm 1\sigma$ are presented in Table \ref{tab:unresolved}. The left column of Figure~\ref{fig:johnSuperPlt} shows a comparison between the observed SED values and the best-fit model SED, as well as the best-fit model image and residual contours for each disk.

\subsubsection{Disks with Resolved Inner Radii}
\label{sec:resolved}

For cases in which both $R_\mathrm{In}$ and $\Delta R$ are at least marginally resolved (HIP 62657, HIP 73145, HIP 79516, and HIP 76310), we fix the full geometry of the disk ($R_\mathrm{In}$, $\Delta R$, $i$, $PA$) to the best fit values from the visibilities only fit. For cases in which $R_\mathrm{In}$ is resolved but $\Delta R$ is not (HIP 79742 and HIP 79977), we set $R_\mathrm{In}$, $i$, and $PA$ to their best fit values and assume that $\Delta R$ is 1/3 the width of the beam. 

The simplest means of fitting the SED involves varying $a$ and $M_\mathrm{Disk}$ with the geometry fixed to the best fit values to the visibilities, but we find that we can not recreate the observed photometry for these six disks, as our models are not bright enough in the mid-IR (raw $\chi^{2}$ in Column \RNum{6}, Table \ref{tab:AIC}). We add an additional, unresolved inner belt with inner edge $R_{In,~Inner~Belt}$, width 1/10 the resolution of the beam, mass $M_{Belt}$, and $\Sigma(r) \propto r^{-1}$. Varying $a$, $M_\mathrm{Disk}$, $R_{In,~Inner~Belt}$, and $M_{Belt}$, we find that we are able to recreate both the SED and the visibilities (raw $\chi^{2}$ in Column \RNum{7}), and that these models are all significantly better than the models without an inner belt at the $> 10\sigma$ level (Column \RNum{8}). Best fit and Median values $\pm 1\sigma$ are presented in Table \ref{tab:resolved}. Since the inner belt is spatially unresolved and does not contribute significant flux to the long-wavelength image, we are unable to determine whether the additional mid-IR flux in fact results from a spatially disparate belt with the same grain size (as we assume in the model), or whether it results from a distinct population of smaller (and therefore hotter) grains that are spatially co-located with the grains in the outer disk.  Due to the large difference in temperature, this latter possibility would require an essentially bimodal grain size distribution.  Since the best-fit characteristic grain size for many of the disks is already comparable to the blowout size, a two-belt scenario rather than a population of grains significantly smaller than the blowout size seems more likely, but we are unable to distinguish conclusively between the two scenarios based on the available data.  

The median and best-fit values for each parameter in the fit are reported in Table~\ref{tab:resolved}.  The right column of Figure~\ref{fig:johnSuperPlt} shows a comparison between the observed SED values and the best-fit model SED, as well as the best-fit model image and residual contours for each disk.  

\section{Discussion}
\label{sec:discussion}

\subsection{Size and Geometry from Visibilities}

Of the disks that are spatially resolved by our observations, half exhibit only spatially resolved outer radii with the inner radius unresolved, while the other half exhibit spatially resolved inner radii with either resolved or unresolved widths.  With typical Briggs-weighted beam sizes of 0.5-1\,arcsec and stellar distances of 85-150\,pc, the range of linear diameters corresponding to the spatial resolution represented in the data varies between $\sim$40 and 150\,AU, allowing us to probe the radial structure of the disk on scales of $\sim$20-75\,AU and larger (due to the nature of the \texttt{CLEAN} deconvolution algorithm, a visibility analysis of the data set is typically sensitive to spatial scales somewhat smaller than those corresponding to the Briggs-weighted beam, which accounts for the smallest upper limits in Tables~\ref{tab:unresolved} and \ref{tab:resolved}).  The ability of the data to measure disk widths is limited by the spatial resolution ($\sim$40-150\,AU), so that we are only able to measure the disk width for disks that are relatively broad in relation to their sizes ($\Delta R / R \sim 1$ and larger).  All six of the disks with unresolved inner radii have measurably broad widths, while of the six disks with resolved inner radii only three have spatially resolved widths.  

\begin{figure}[ht!]
\centering
\includegraphics[width=0.5\textwidth]{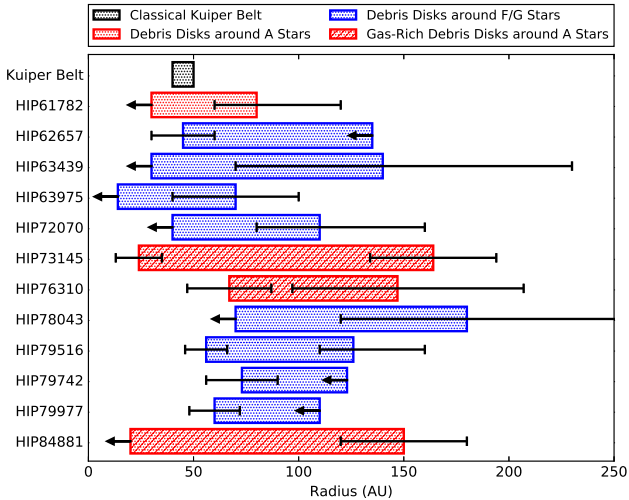}
\caption{
Graphical representation of the best-fit inner and outer radii of disks in the sample around A stars (red) and F/G stars (blue), compared with the dimensions of the classical Kuiper belt (black).  Upper limits on inner radii are indicated by left-pointing arrows; upper limits on the disk width are indicated by left-pointing arrows that extend from 1$\sigma$ beyond the (resolved) inner radius. 
}
\label{fig:rin_rout}
\end{figure}

Constraints on the inner and outer radii of the 12 spatially resolved disks in the sample are summarized and compared with the classical Kuiper belt in Figure~\ref{fig:rin_rout}.  Compared to the classical Kuiper belt, which has an inner radius of 40\,AU and an outer radius of 50\,AU \citep[][and references therein]{bar08}, a majority of the debris disks in our sample are noticeably more radially broad: nine of the 12 disks have spatially resolved widths of 70\,AU or more, while the classical Kuiper belt has a far narrower radial width of only 10\,AU (which would be unresolved by our data).  It is possible that the resolved disks in our sample are more analogous to the scattered component of the Kuiper belt, which extends for a width of hundreds of AU beyond its 40\,AU inner radius. While events like those thought to be responsible for creating the scattered Kuiper belt in our own solar system are thought to be rare in mature systems \citep{boo09}, the systems in our sample were selected for their relatively young ($\sim$10\,Myr) ages, which may be responsible for the prevalence of broad disks in our sample.  It is also possible that higher resolution observations would reveal a series of narrow belts instead of a single broad belt, or a more complicated dust distribution of gaps and/or regions of enhanced density superimposed on a power law background \citep[see, e.g., ][]{ric15,hug16}.

There is also no obvious trend of disk geometry with stellar type; the four A stars with spatially resolved disks span the full range of inner radii in the sample.  All three of the disks with unresolved widths are around F stars, but since F stars make up the majority of stars in the sample with spatially resolved disks, this could easily be a chance occurrence.  These results are consistent with those of \citet{paw14}, who find that disk sizes are independent of stellar luminosity.  Our results therefore reinforce their conclusion that ice line locations do not play an important role in determining the location of dust rings in debris disks, which would otherwise predict a relationship between the disk inner radius and the stellar luminosity that we do not observe \citep[see also][]{bal13}.  We are not able to investigate the weak correlation between disk size and stellar age suggested by \citet{eir13}, since our objects were selected to have similar $\sim$10\,Myr ages, although we note that the radii of the spatially resolved disks in our sample are as large as the oldest disks in their sample -- far larger than would be expected from an extrapolation of their observed trend to these younger ages as illustrated in the bottom center panel of their Fig. 11 (although the larger disk sizes we observe may be due to our selection bias towards brighter sources). 

Due to the relatively low spatial resolution of the data and correspondingly large uncertainties on disk dimensions, nearly all of the disks have dimensions consistent with those of the Kuiper belt at the 3$\sigma$ level (including those that are spatially unresolved by our observations).  Four of the disks in the sample (HIP 63439, HIP 61782, HIP 63975, and HIP 84881) have inner radii significantly smaller than that of the Kuiper belt, while only two disks (HIP 84881 and HIP 73145) have outer radii that are larger than that of the Kuiper belt at the $>3\sigma$ level.  Interestingly, these disks with significantly larger outer radii comprise two of the three A star disks in the sample that also host a significant amount of molecular gas \citep{hal16}.  The third, HIP 76310, also hosts an extended, spatially resolved debris disk, but it only differs in its dimensions from the classical Kuiper belt at the $2\sigma$ level.  

The three CO-rich disks comprise three of the four disks in the sample with the largest outer radii; the fourth, HIP 78043, also has one of the largest uncertainties in outer radius of all of the disks in the sample due to the low signal-to-noise ratio of the detection.  Since the spatially unresolved sources are all smaller than these resolved disks, we can make the stronger observation that the CO-rich sources comprise three of the four largest disks in the full sample of 20 detected debris disks.  This trend hints that gas-rich debris disks may be more spatially extended on average than their gas-poor counterparts around stars of similar ages.  This may be expected if the CO is optically thick, since the flux in that case would be the temperature times the disk surface area, which would bias the sample towards detections in radially larger disks (although the sharp contrast in flux between CO detections and non-detections discussed in Section~\ref{sec:co21} makes this unlikely).  It is also consistent with a scenario in which as a dust disk becomes optically thin in the presence of gas, the orbits modified by the outward force of radiation pressure begin to experience a tail wind from the gas and can be ejected to larger orbits, potentially even exceeding the radius of the gas disk \citep{tak01}.  Such a scenario is consistent with the recent observation that CO emission in the HD~141569 disk is confined within the spectacular optically thin dust rings imaged by {\it HST} \citep{fla16}.  

A perennial question in studies of debris disks around nearby stars is the degree of axisymmetry of the disk.  Optical surveys \citep[e.g.][]{sch14} find that highly asymmetric disks and out-of-plane features and substructures are commonly observed in high-resolution images of scattered light from small grains in the outer disk.  Many of these features have been attributed to the presence of underlying planetary systems, or interactions between the disk and the interstellar medium through which it is passing.  While the spatial resolution and sensitivity of our data are far lower than the optical surveys that reveal these features, it is worth noting that we are able to reproduce all of the observed data without a need for asymmetries or clumpy structure in these debris disks.  In fact, the only debris disk that has yet been demonstrated to exhibit statistically significant departures from axisymmetry when observed with millimeter interferometry is the disk around $\beta$ Pictoris \citep{den14}, although it seems likely (at the 3$\sigma$ level) that HD 15115 is also asymmetric \citep{mac15a}.  Given the current limitations in sensitivity and angular resolution, we would only be able to detect very large asymmetries in the disks in our sample (with flux differences of order 50-100\% between synthesized beams), and it is certainly possible that future studies will reveal more subtle features like warps, eccentricities, or subtle density contrasts.  These results are consistent with those of a similar set of observations of debris disks around Solar-type stars collected and interpreted by \citet{stee15}.

\subsection{Disk Masses and Grain Sizes from SED and Visibilities}

\begin{figure*}[ht!]
\centering
\includegraphics[width=1.0\textwidth]{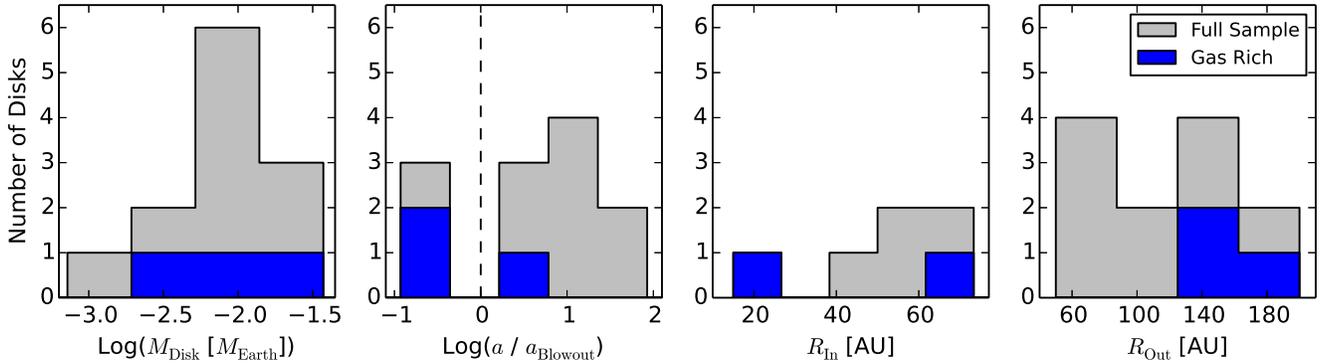}
\caption{
Histograms of the best-fit parameters, including disk mass (far left), grain size relative to the blowout size (left center), inner radius in the subset of six disks for which it was resolved (right center), and outer radius for the spatially resolved disks (far right).  The gray shaded regions represent histograms for the full sample of 12 spatially resolved sources, while the dark blue regions represent the properties of only the three gas-rich disks (HIP~84881, HIP~73145, and HIP~76310).  For the outer radius histogram, for disks with spatially resolved inner radii but upper limits on disk width, we have set the width of the disk to be 1/3 the $3\sigma$ upper limit on disk width to provide an approximate estimate of the location of the outer radius.  The vertical dashed line in the grain size panel represents the location of a hypothetical disk in which the characteristic grain size is equal to the blowout size for a given stellar mass and luminosity.  
}
\label{fig:hist}
\end{figure*}

Figure~\ref{fig:hist} presents histograms of the four primary diagnostics of the disk surface density structure and dust grain sizes that we were able to measure for our sample (disk mass $M_\mathrm{Disk}$, characteristic grain size $a$ relative to blowout size $a_\mathrm{Blowout}$, inner radius $R_\mathrm{In}$ for the sample of six objects for which it is spatially resolved, and the outer radius $R_\mathrm{Out}$).  

Only three of the 12 disks in the spatially resolved sample exhibit characteristic grain sizes smaller than the theoretical blowout size for the corresponding stellar mass and luminosity.  Two of those three disks are gas-rich debris disks.  With such a small sample size, and particularly with such a small number of gas-rich disks, the association may well be by chance; however, it is perhaps plausible to imagine that gas-rich debris disks may be more likely to hold onto their small grains due to drag forces from the gas, or that they may be more rich in small grains due to recent collisions that have given rise to the excess gas in the system as well as a cascade of small dust grains.  Previous surveys have found that although dust temperatures in Kuiper Belt-like debris disks tend to be higher around more luminous stars, the dust temperature {\it relative to the blackbody equilibrium temperature} is lower for disks around more luminous stars \citep{bal13,boo13,eir13,che14,paw14} -- a trend that we do not recover in our sample (in the context of our model, such a trend would manifest as a correlation between effective grain size and stellar luminosity).  However, among our sample size of 12 spatially resolved disks, at least three and possibly all four of the A stars in the sample are gas-rich.  It is possible that the trend previously identified at mid- and far-IR wavelengths of finding larger, colder grains in disks around more luminous stars does not hold for gas-rich debris disks.  

Since the sample was selected according for large infrared excess, it is perhaps not surprising that the disk masses in the spatially resolved sample are comparable to the brightest debris disks detected around other nearby main sequence stars \citep[e.g.][]{roc09,thu14}.  Within this biased sample of objects, there is no obvious trend relating the debris disk mass to the spectral type of the central star or the presence of substantial quantities of gas in the disk.  This is interesting because the dust mass in a debris disk might reasonably be expected to be related to the collision rate between planetesimals, and if the gas is second-generation then it would require a large collision rate \citep[the equivalent of vaporizing several Hale-Bopp-size objects per minute, according to estimates based on similar gas-rich disks in][]{kos13,den14} to sustain the quantities of molecular gas that we observe in such close proximity to the photodissociating radiation from the central A star.  

\section{Summary and Conclusions}
\label{sec:conclusions}

We have presented 1240\,$\mu$m observations of a sample of 24 disks imaged at 0.5-1\,arcsec resolution with the ALMA interferometer.  20 of the 24 disks are detected in continuum emission at the $>3\sigma$ level; of these disks, 12 are spatially resolved by our observations.  Of the spatially resolved disks, analysis of the ALMA visibilities reveals disk emission over a wide range of radii between $\sim$50 and 200\,AU.  The disk geometry is at least broadly consistent with that of the Kuiper belt in terms of the disk radii, although radially broad disks appear to be more common than narrow disks like the classical Kuiper belt.  Given the limited spatial resolution of the data, we cannot yet determine the details of the surface density profile, including whether the dust is indeed distributed in a single broad belt or whether it is instead concentrated into several narrow rings.  Combined with the overall larger masses of these debris disks, they appear to be analogous to scaled-up versions of the Kuiper belt, perhaps more similar to the scattered belt than the classical belt.  

Combining the geometrical constraints from the visibilities with the temperature information encoded in the spectral energy distribution, we also fit for the characteristic grain size and mass of the disk.  For all six disks with spatially resolved inner radii, we require the presence of an additional, warmer belt of dust that can reproduce the mid-IR flux excess present in the SED without contributing significant emission to the ALMA image at millimeter wavelengths.  In the absence of infrared images of these disks, we cannot distinguish between a population of dust grains smaller than the blowout size in the outer disk or a spatially distinct population of dust grains at the characteristic grain size concentrated in an asteroid belt near the star; we model the SED assuming the latter, but cannot rule out the former given the currently available data.  These results are consistent with those of previous surveys that have found that multi-temperature disk components are frequently required to reproduce observed SEDs and disk images \citep[e.g.,][]{su13,paw14,stee15}.  

The three strongly CO-rich debris disks in the sample represent three of the four disks with the largest outer radii and two of the three disks with characteristic grain sizes less than the blowout size.  These results provide suggestive, although not conclusive, evidence that gas-rich disks may be preferentially more extended and contain smaller grains than their gas-poor counterparts.  Despite the presence of other resolved A star debris disks in the sample, there appears to be no correlation between the presence of substantial quantities of molecular gas and the dust mass inferred from continuum emission.  If the gas is of second-generation origin, this result is puzzling because it violates the expectation that a higher collision rate between icy KBO-like planetesimals is responsible for the larger CO mass visible in the system, since a correspondingly larger dust mass would also be expected.  

\acknowledgments
The authors thank Angelo Ricarte for his contributions to the code base and helpful comments, and the anonymous referee for a careful commentary that improved the paper.  J.~L.-S. and A.~M.~H. gratefully acknowledge support from NSF grant AST-1412647.  J.~L.-S. was also supported in part by a NASA CT Space Grant Directed Campus Scholarship.   We acknowledge Wesleyan University for time on its high performance computing cluster, supported by the NSF under grant number CNS-0619508.  This work makes use of the following ALMA data: ADS/JAO.ALMA\#2012.1.00688.S.  ALMA is a partnership of ESO (representing its member states), NSF (USA), and NINS (Japan), together with NRC (Canada) and NSC and ASIAA (Taiwan), in cooperation with the Republic of Chile.  The Joint ALMA Observatory is operated by ESO, AUI/NRAO, and NAOJ.  The National Radio Astronomy Observatory is a facility of the National Science Foundation operated under cooperative agreement by Associated Universities, Inc.  This research has made use of NASA's Astrophysics Data System Bibliographic Services, as well as Astropy, a community-developed core Python package for Astronomy \citep{astropy}.  

\nocite{*}
\bibliographystyle{apj}
\bibliography{ms}

\end{document}